\documentclass[notitlepage]{article}

\usepackage{authblk}
\usepackage{graphicx}
\usepackage{subcaption}
\usepackage{amsmath}
\usepackage{multirow}
\usepackage{booktabs}
\usepackage{topcapt}
\usepackage{xcolor}
\usepackage{amssymb}
\usepackage{times}
\usepackage{comment} 

\usepackage{caption}
\usepackage{subcaption}

\usepackage[caption=false]{subfig}
\usepackage{pgfplots}
\usepackage{pgfplotstable}
\usepackage{adjustbox}
\usepgfplotslibrary{groupplots}
\usepackage{tikz}
\usetikzlibrary{arrows,decorations.markings}
\usetikzlibrary{shapes.arrows}
\usetikzlibrary{patterns}
\usetikzlibrary{matrix}
\tikzstyle{internalnode} = [circle, draw, fill, minimum size=1.7mm,inner sep=0pt,outer sep=0pt]

\usepackage{titling}

\newcommand{\myparagraph}[1]{\medskip\noindent\textbf{#1}\quad}

\newcommand\eg{\emph{e.g.}}
\newcommand\ie{\emph{i.e.}}
\newcommand\etal{\emph{et al.}}

\newcommand{\nop}[1]{}

\newenvironment{customlegend}[1][]{%
    \begingroup
    \csname pgfplots@init@cleared@structures\endcsname
    \pgfplotsset{#1}%
}{%
    \csname pgfplots@createlegend\endcsname
    \endgroup
}%

\def\addlegendimage{\csname pgfplots@addlegendimage\endcsname}

\begin{document}

\title{The Infinity Mirror Test for Analyzing the Robustness of Graph Generators}
\author{Salvador Aguinaga}
\author{Tim Weninger}
\affil{Department of Computer Science and Engineering\\University of Notre Dame}
\date{}                     
\setcounter{Maxaffil}{0}
\renewcommand\Affilfont{\itshape\small}
    \maketitle
    \begin{abstract}
Graph generators learn a model from a source graph in order to generate a new graph that has many of the same properties. The learned models each have implicit and explicit biases built in, and its important to understand the assumptions that are made when generating a new graph. Of course, the differences between the new graph and the original graph, as compared by any number of graph properties, are important indicators of the biases inherent in any modelling task. But these critical differences are subtle and not immediately apparent using standard performance metrics. Therefore, we introduce the infinity mirror test for the analysis of graph generator performance and robustness. This stress test operates by repeatedly, recursively fitting a model to itself. A perfect graph generator would have no deviation from the original or ideal graph, however the implicit biases and assumptions that are cooked into the various models are exaggerated by the infinity mirror test allowing for new insights that were not available before. We show, via hundreds of experiments on 6 real world graphs, that several common graph generators do degenerate in interesting and informative ways. We believe that the observed degenerative patterns are clues to future development of better graph models.
\end{abstract}

\section{Introduction}

Teasing out interesting relationships buried within volumes of data is one of the most basic challenges in data science research. When this data is viewed as an information network, the standard approach is to treat the network as a graph with some number of nodes and edges. Increasingly, researchers and practitioners are interested in understanding how individual pieces of information are organized and interact in order to discover the fundamental principles that underlie a physical or social phenomena.

With this motivation, researchers have developed a suite of graph generation techniques that learn a model of a network in order to extrapolate, generalize or otherwise gain a deeper understanding of the data set. Early graph generators like the Erd\H{o}s-R\'{e}nyi, Watts-Strogatz, and Barabasi-Albert models produce random graphs, small world graphs, and scale free graphs respectively. Although they are used to generate graphs given some hand-picked parameters, they do not learn a model from any observed real-world network.

We focus instead on graph model inducers, which take some observed network $G$, learn a model $\Theta$ and produce a new graph $G^\prime$. These types of graph generators include the Kronecker Model, Chung-Lu Model, Exponential Random Graph Model (ERGM) and Block Two-Level Erd\H{o}s-R\'{e}nyi Model (BTER), and others.

The performance of a graph generator can be judged based on how well the new graph matches certain topological characteristics of the original graph. Unfortunately small perturbations caused by the implicit and inherent biases of each type of model may not be immediately visible using existing performance metrics.

In the present work, we address this problem by characterizing the \textit{robustness} of a graph generator via a new metric we call the \textit{infinity mirror test}. The ``infinity mirror'' gets its name from the novelty item with a pair of mirrors, set up so as to create a series of smaller and smaller reflections that appear to taper to an infinite distance. The motivating question here is to see if a generated graph $G^\prime$ holds sufficient information to be used as reference. Although a comparison between $G$ and $G^\prime$ may show accurate results, the model's biases only become apparent after recursive application of the model onto itself. 

The details of the method are discussed later, but, simply put, the infinity mirror tests the robustness of a graph generator because errors (or biases) in the model are propagated forward depending on their centrality and severity. A robust graph generator, without severe biases or errors, should remain stable after several recurrences. However, a non-robust model will quickly degenerate, and the manner in which the model degenerates reveals the model-biases that were hidden before. 

\section{Graph Generators}

Several graph generators have been developed for the tasks outlined above. We describe some of them here.

\myparagraph{Kronecker Graph}
Kronecker graphs operate by learning a $2\times 2$ initiator matrix $K_1$ of the form

$$K_1 = 
\begin{bmatrix}
    k_{1}       & k_{2} \\
    k_{3}       & k_{4} \\
\end{bmatrix}
$$

and then performing a recursive multiplication of that initiator matrix in order to create a probability matrix $P_{Kron}$ from which we can stochastically pick edges to create $G^\prime$. Because of the recursive multiplication, the Kronecker product only creates graphs where the number of nodes is an exponential factor of 2, \ie, $2^x$~\cite{Leskovec2010kronecker}. 

The initiator matrix can be learned quickly, and the final graph shares many similarities with the original graph making the Kronecker graph model a natural fit for many graph modelling tasks. 

\myparagraph{Chung-Lu Models} The Chung-Lu Graph Model takes, as input, some empirical (or desired) degree distribution and generates a new graph of the similar degree distribution and size~\cite{ChungLu2002connected,chung2002average}. An optimized version called Fast Chung-Lu (FCL) was developed analogous to how the Kronecker model samples its final graph. Suppose we are given sequences of $n$-degrees $d_1$, $d_2$, $\ldots$ $d_n$ where $\sum_{i}{d_i} = 2m$. We can create a probability matrix $P_{FCL}$ where the edge $e_{ij}$ has a probability $d_i d_j/{m^2}$~\cite{pinar2011similarity}.

On average, the Chung-Lu model is shown to preserve the degree distribution of the input graph. However, on many graphs, the clustering coefficients and assortativity metrics of the output graphs do not match the original graph. Extensions of the Chung-Lu (CL) model, such as Transitive CL (TCL)~\cite{pfeiffer2012fast}, Binning CL (BCL)~\cite{mussmann2014assortativity} and Block Two-Level Erd\H{o}s-R\'{e}nyi Model (BTER)~\cite{kolda2014scalable}, have been developed to further improve performance.

\myparagraph{Exponential Random Graph}
Exponential Random Graph Models (ERGMs) are a class of probabilistic models used to directly describe several structural features of a graph~\cite{Robins2007}. Although ERGMs have been shown to model the degree distributions and other graph properties of small graphs, they simply do not scale to graphs of even moderate size. As a result we cannot include ERGM in the present work.

Existing approaches to graph modelling and generation perform well in certain instances, but each have their drawbacks. The Kronecker Model, for example, can only represent graphs with a power law degree distribution. Both Kronecker and the Chung-Lu models ignore local subnetwork properties, giving rise to more complex models like Transitive Chung-Lu for better clustering coefficient results~\cite{pfeiffer2012fast} or Chung-Lu with Binning for better assortativity results~\cite{mussmann2014assortativity,Mussmann:2015}. Exponential Random Graph Models (ERGMs) take into consideration the local substructures of a given graph. However, each substructure in an ERGM must be pre-identified by hand, and the complexity of the model increases (at least) quadratically as the size of the graph grows.

\section{Infinity Mirror Test}

We characterize the \textit{robustness} of a graph generator by its ability to repeatedly learn and regenerate the same model. A perfect, lossless model (\eg, $\Theta = G$) would generate $G^\prime$ as an isomorphic copy of the original graph. If we were to again apply the perfect model on the isomorphic $G^\prime$, we would again generate an isomorphic copy of the graph. On the other hand, a non-robust graph generator may generate a $G^\prime$ that is dissimilar from $G$; if we were to learn a new model from $G^\prime$ and create a second-level graph, we would expect this second graph to exacerbate the errors (the biases)  that the first graph made and be even less similar to $G$. A third, fourth, fifth, etc. application of the model will cause the initial errors to accumulate and cause cascading effects in each successive layer.

Colored by this perspective, the robustness of a graph generator is defined by its ability to maintain its topological properties as it is recursively applied. To that end, this paper presents the infinity mirror test. In this test, we repeatedly learn a model from a graph generated by the an earlier version of the same model. 

Starting with some real world network $G$, a graph generator learns a model $\Theta_1$ (where the subscript $\cdot_1$ represents the first recurrence) and generates a new graph $G^{\prime}_1$. At this point, current works typically overlay graph properties like degree distribution, assortativity, etc. to see how well $G$ matches $G^{\prime}_1$. We go a step further and ask if the new graph $G^{\prime}_1$ holds sufficient information to be used as reference itself. So, from $G^{\prime}_1$ we learn a new model $\Theta_2$ in order to generate a second-level graph $G^{\prime}_2$. We repeat this recursive ``learn a model from the model''-process $k$ times, and compare $G^{\prime}_k$ with the original graph.

\begin{figure}
\centering
\begin{adjustbox}{width=.78\textwidth}
    \input{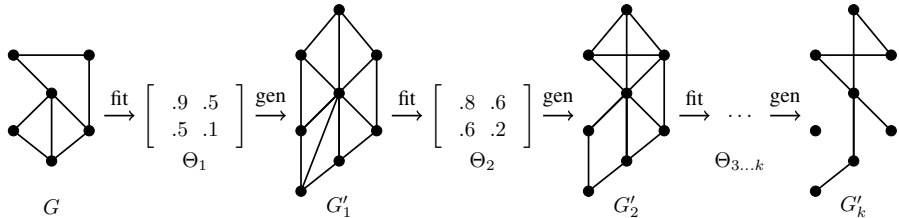}
\end{adjustbox}
\caption{Example infinity mirror test on the Kronecker model. This test recursively learns a model and generates graphs. Although not apparent in $G^\prime_1$, this example shows a particular type of degeneration where the model loses edges. }
\label{fig:mirror}
\end{figure}

Figure~\ref{fig:mirror} shows an example of the infinity mirror test for the Kronecker model. In this example some real world graph $G$ is provided by the user. From $G$ a model $\Theta_1$ is fit, which is used to generate a new graph $G^{\prime}_1$. Of course, $G^{\prime}_1$ is only an approximation of $G$ and is therefore slightly different. In the second recurrence a new model $\Theta_2$ is fit from $G^{\prime}_1$ and used to generate a new graph $G^{\prime}_2$. This continues recursively $k$ times.

With the infinity mirror test, our hypothetical, perfect model is perfectly robust and immune to error. A hypothetical ``bad'' model would quickly degenerate into an unrecognizable graph after only a few recurrences. Despite their accurate performance, existing models are far from perfect. We expect to see that all models degenerate as the number of recurrences grow. The question is: how quickly do the models degenerate and how bad do the graphs become?

\section{Experiments}

In order to get a holistic and varied view of the robustness of various graph generators, we consider real-world networks that exhibit properties that are both common to many networks across different fields, but also have certain distinctive properties.

\begin{table}[b]
\centering
\caption{Real networks}
\begin{tabular}{r|rr}
  \textbf{Dataset Name} & \textbf{Nodes} & \textbf{Edges} \\\hline
  C. elegans neural (male) & 269 & 2,965\\
  Power grid & 4,941 &6,594\\
  ArXiv GR-QC & 5,242 & 14,496\\
  Internet Routers & 6,474 & 13,895\\
  Enron Emails & 36,692 & 183,831 \\
  DBLP & 317,080 & 1,049,866\\
\end{tabular}
\label{tab:realnets}
\vspace{-.1cm}
\end{table}

The six real world networks considered in this paper are described in Table.~\ref{tab:realnets}. The networks vary in their number of vertices and edges as indicated, but also vary in clustering coefficient, degree distribution and many other graph properties. Specifically, C. elegans is the neural network of the roundworm of the named species~\cite{Jarrell437}; the Power grid graph is the connectivity of the power grid in the Western United States~\cite{Watts1998}; the Enron graph is the email correspondence graph of the now defunct Enron corporation~\cite{klimt2004introducing}; the ArXiv GR-QC graph is the co-authorship graph extracted from the General Relativity and Quantum Cosmology section of ArXiv; the Internet router graph is created from traffic flows through Internet peers; and, finally, DBLP is the co-authorship graph from the DBLP dataset. All datasets were downloaded from the SNAP and KONECT dataset repositories.

On each of the six real world graphs, we recursively applied the Kronecker, Block Two-Level Erdos-Renyi (BTER), Exponential Random Graph (ERGM) and Chung-Lu (CL) models to a depth of $k$=10.

\begin{figure}[t]
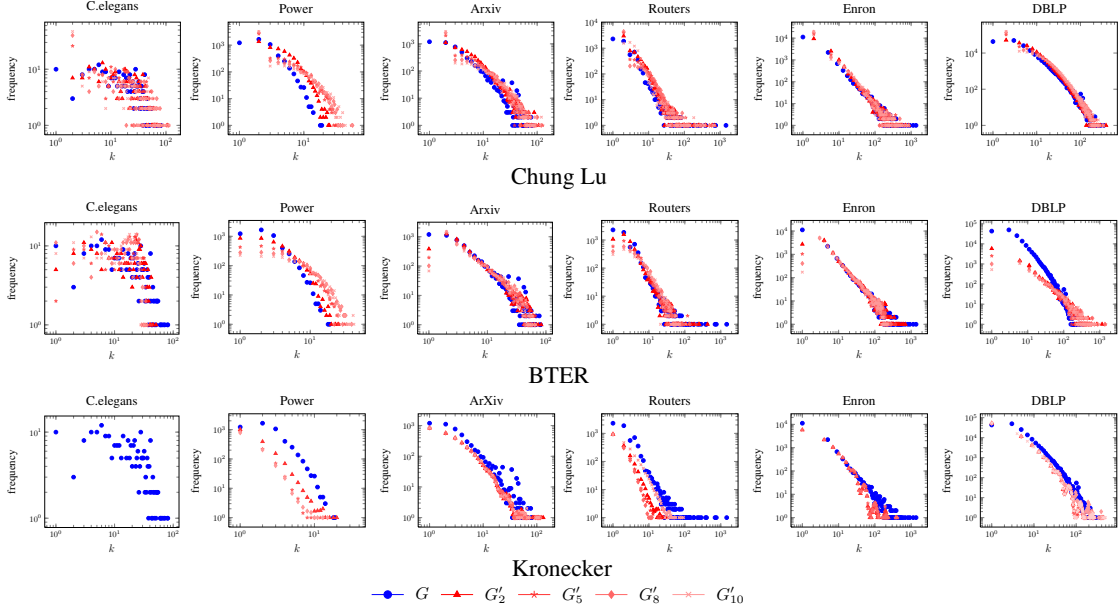

\centering
\begin{minipage}[b]{0.16\linewidth}
    \resizebox{\textwidth}{!}{
        \input{./nfigs/fcl/celegans_degree_fcl}
    }
\end{minipage}%
\begin{minipage}[b]{0.16\linewidth}
    \resizebox{\textwidth}{!}{
        \input{./nfigs/fcl/power_degree_fcl}
    }
\end{minipage}    
\begin{minipage}[b]{0.16\linewidth}
    \resizebox{\textwidth}{!}{
        \input{./nfigs/fcl/arxiv_degree_fcl}
    }
\end{minipage}%
\begin{minipage}[b]{0.16\linewidth}
    \resizebox{\textwidth}{!}{
        \input{./nfigs/fcl/router_degree_fcl}
    }
\end{minipage}
\begin{minipage}[b]{0.16\linewidth}
    \resizebox{\textwidth}{!}{
        \input{./nfigs/fcl/enron_degree_fcl}
    }
\end{minipage}
\begin{minipage}[b]{0.16\linewidth}
    \resizebox{\textwidth}{!}{
        \input{./nfigs/fcl/dblp_degree_fcl}
    }
\end{minipage}
\\
\vspace{-1mm}
\small{Chung Lu}
\vspace{1mm}
\\
\begin{minipage}[b]{0.16\linewidth}
    \resizebox{\textwidth}{!}{
        \input{./nfigs/bter/celegans_degree_bter}
    }
\end{minipage}%
\begin{minipage}[b]{0.16\linewidth}
    \resizebox{\textwidth}{!}{
        \input{./nfigs/bter/power_degree_bter}
    }
\end{minipage}    
\begin{minipage}[b]{0.16\linewidth}
    \resizebox{\textwidth}{!}{
        \input{./nfigs/bter/arxiv_degree_bter}
    }
\end{minipage}%
\begin{minipage}[b]{0.16\linewidth}
    \resizebox{\textwidth}{!}{
        \input{./nfigs/bter/router_degree_bter}
    }
\end{minipage}
\begin{minipage}[b]{0.16\linewidth}
    \resizebox{\textwidth}{!}{
        \input{./nfigs/bter/enron_degree_bter}
    }
\end{minipage}
\begin{minipage}[b]{0.16\linewidth}
    \resizebox{\textwidth}{!}{
        \input{./nfigs/bter/dblp_degree_bter}
    }
\end{minipage}
\\
\vspace{-1mm}
\small{BTER}
\vspace{1mm}
\\
\begin{minipage}[b]{0.16\linewidth}
    \resizebox{\textwidth}{!}{
        \input{./nfigs/kron/celegans_degree_kron}
    }
\end{minipage}%
\begin{minipage}[b]{0.16\linewidth}
    \resizebox{\textwidth}{!}{
        \input{./nfigs/kron/power_degree_kron}
    }
\end{minipage}    
\begin{minipage}[b]{0.16\linewidth}
    \resizebox{\textwidth}{!}{
        \input{./nfigs/kron/arxiv_degree_kron}
    }
\end{minipage}%
\begin{minipage}[b]{0.16\linewidth}
    \resizebox{\textwidth}{!}{
        \input{./nfigs/kron/router_degree_kron}
    }
\end{minipage}
\begin{minipage}[b]{0.16\linewidth}
    \resizebox{\textwidth}{!}{
        \input{./nfigs/kron/enron_degree_kron}
    }
\end{minipage}
\begin{minipage}[b]{0.16\linewidth}
    \resizebox{\textwidth}{!}{
        \input{./nfigs/kron/dblp_degree_kron}
    }
\end{minipage}
\\
\vspace{-1mm}
\small{Kronecker}
\\
\begin{minipage}[b]{0.35\linewidth}
    \resizebox{\textwidth}{!}{
        \input{./figs/legend}
    }
\end{minipage}%
\caption{Degree distribution. $G$ shown in blue. $G^\prime_2$, $G^\prime_5$, $G^\prime_8$ and $G^\prime_{10}$ are shown in lighter and lighter shades of red. Degeneration is observed when recurrences increasingly deviate from $G$.}
\label{fig:deg_results}
\end{figure}
\begin{figure}[t]
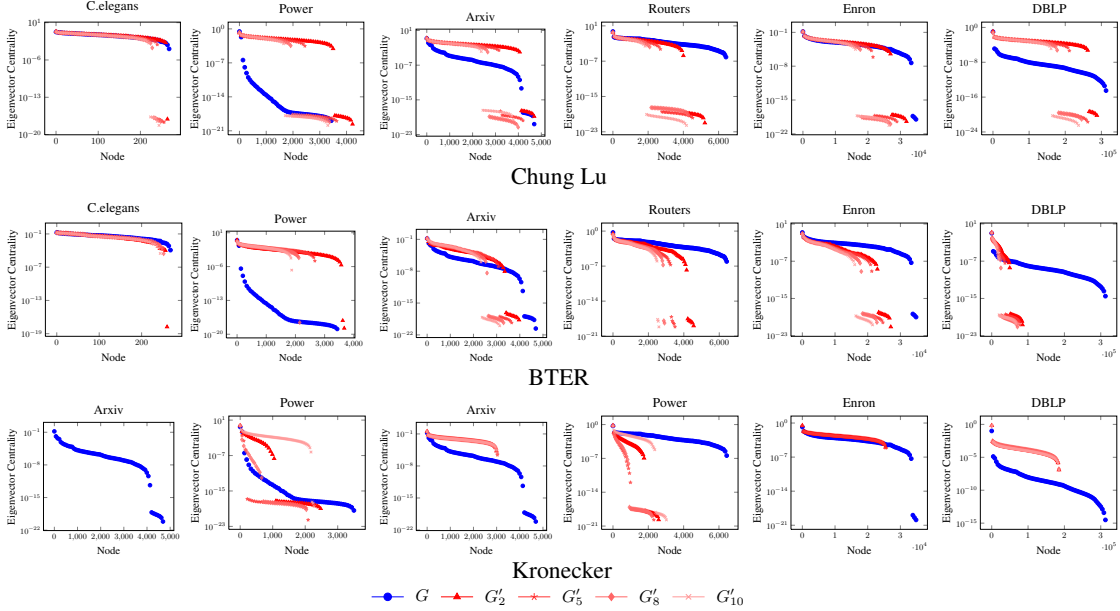

\centering
\begin{minipage}[b]{0.16\linewidth}
    \resizebox{\textwidth}{!}{
        \input{./nfigs/fcl/celegans_eig_fcl}
    }
\end{minipage}%
\begin{minipage}[b]{0.16\linewidth}
    \resizebox{\textwidth}{!}{
        \input{./nfigs/fcl/power_eig_fcl}
    }
\end{minipage}    
\begin{minipage}[b]{0.16\linewidth}
    \resizebox{\textwidth}{!}{
        \input{./nfigs/fcl/arxiv_eig_fcl}
    }
\end{minipage}%
\begin{minipage}[b]{0.16\linewidth}
    \resizebox{\textwidth}{!}{
        \input{./nfigs/fcl/router_eig_fcl}
    }
\end{minipage}
\begin{minipage}[b]{0.16\linewidth}
    \resizebox{\textwidth}{!}{
        \input{./nfigs/fcl/enron_eig_fcl}
    }
\end{minipage}
\begin{minipage}[b]{0.16\linewidth}
    \resizebox{\textwidth}{!}{
        \input{./nfigs/fcl/dblp_eig_fcl}
    }
\end{minipage}
\\
\vspace{-1mm}
\small{Chung Lu}
\vspace{1mm}
\\
\begin{minipage}[b]{0.16\linewidth}
    \resizebox{\textwidth}{!}{
        \input{./nfigs/bter/celegans_eig_bter}
    }
\end{minipage}%
\begin{minipage}[b]{0.16\linewidth}
    \resizebox{\textwidth}{!}{
        \input{./nfigs/bter/power_eig_bter}
    }
\end{minipage}    
\begin{minipage}[b]{0.16\linewidth}
    \resizebox{\textwidth}{!}{
        \input{./nfigs/bter/arxiv_eig_bter}
    }
\end{minipage}%
\begin{minipage}[b]{0.16\linewidth}
    \resizebox{\textwidth}{!}{
        \input{./nfigs/bter/router_eig_bter}
    }
\end{minipage}
\begin{minipage}[b]{0.16\linewidth}
    \resizebox{\textwidth}{!}{
        \input{./nfigs/bter/enron_eig_bter}
    }
\end{minipage}
\begin{minipage}[b]{0.16\linewidth}
    \resizebox{\textwidth}{!}{
        \input{./nfigs/bter/dblp_eig_bter}
    }
\end{minipage}
\\
\vspace{-1mm}
\small{BTER}
\vspace{1mm}
\\
\begin{minipage}[b]{0.16\linewidth}
    \resizebox{\textwidth}{!}{
        \input{./nfigs/kron/celegans_eig_kron}
    }
\end{minipage}%
\begin{minipage}[b]{0.16\linewidth}
    \resizebox{\textwidth}{!}{
        \input{./nfigs/kron/power_eig_kron}
    }
\end{minipage}    
\begin{minipage}[b]{0.16\linewidth}
    \resizebox{\textwidth}{!}{
        \input{./nfigs/kron/arxiv_eig_kron}
    }
\end{minipage}%
\begin{minipage}[b]{0.16\linewidth}
    \resizebox{\textwidth}{!}{
        \input{./nfigs/kron/router_eig_kron}
    }
\end{minipage}
\begin{minipage}[b]{0.16\linewidth}
    \resizebox{\textwidth}{!}{
        \input{./nfigs/kron/enron_eig_kron}
    }
\end{minipage}
\begin{minipage}[b]{0.16\linewidth}
    \resizebox{\textwidth}{!}{
        \input{./nfigs/kron/dblp_eig_kron}
    }
\end{minipage}
\\
\vspace{-1mm}
\small{Kronecker}
\\
\begin{minipage}[b]{0.35\linewidth}
    \resizebox{\textwidth}{!}{
        \input{./figs/legend}
    }
\end{minipage}%
\caption{Eigenvector centrality. $G$ shown in blue. Results for recurrences $G^\prime_2$, $G^\prime_5$, $G^\prime_8$ and $G^\prime_{10}$ in lighter and lighter shades of red showing eigenvector centrality for each network node. Degeneration is shown by increasing deviation from $G$'s eigenvector centrality signature.}
\label{fig:eig_results}
\end{figure}
\begin{figure}[t]
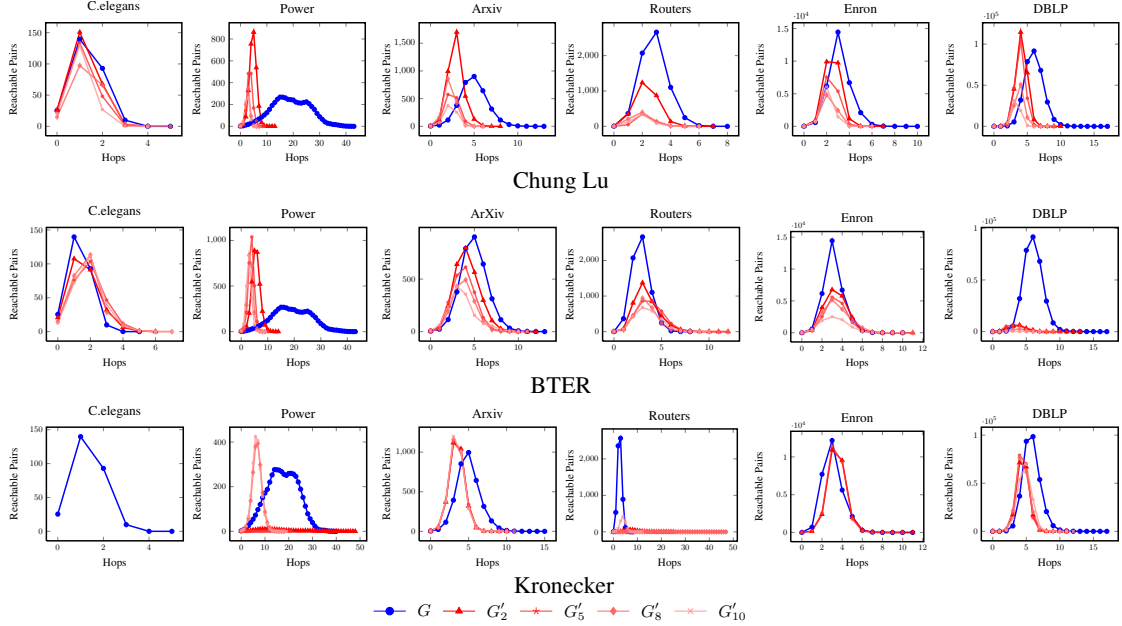

\centering
\begin{minipage}[b]{0.16\linewidth}
    \resizebox{\textwidth}{!}{
        \input{./nfigs/fcl/celegans_hopplot_fcl}
    }
\end{minipage}%
\begin{minipage}[b]{0.16\linewidth}
    \resizebox{\textwidth}{!}{
        \input{./nfigs/fcl/power_hopplot_fcl}
    }
\end{minipage}    
\begin{minipage}[b]{0.16\linewidth}
    \resizebox{\textwidth}{!}{
        \input{./nfigs/fcl/arxiv_hopplot_fcl}
    }
\end{minipage}%
\begin{minipage}[b]{0.16\linewidth}
    \resizebox{\textwidth}{!}{
        \input{./nfigs/fcl/router_hopplot_fcl}
    }
\end{minipage}
\begin{minipage}[b]{0.16\linewidth}
    \resizebox{\textwidth}{!}{
        \input{./nfigs/fcl/enron_hopplot_fcl}
    }
\end{minipage}
\begin{minipage}[b]{0.16\linewidth}
    \resizebox{\textwidth}{!}{
        \input{./nfigs/fcl/dblp_hopplot_fcl}
    }
\end{minipage}
\\
\vspace{-1mm}
\small{Chung Lu}
\vspace{1mm}
\\

\begin{minipage}[b]{0.16\linewidth}
    \resizebox{\textwidth}{!}{
        \input{./nfigs/bter/celegans_hopplot_bter}
    }
\end{minipage}%
\begin{minipage}[b]{0.16\linewidth}
    \resizebox{\textwidth}{!}{
        \input{./nfigs/bter/power_hopplot_bter}
    }
\end{minipage}    
\begin{minipage}[b]{0.16\linewidth}
    \resizebox{\textwidth}{!}{
        \input{./nfigs/bter/arxiv_hopplot_bter}
    }
\end{minipage}%
\begin{minipage}[b]{0.16\linewidth}
    \resizebox{\textwidth}{!}{
        \input{./nfigs/bter/router_hopplot_bter}
    }
\end{minipage}
\begin{minipage}[b]{0.16\linewidth}
    \resizebox{\textwidth}{!}{
        \input{./nfigs/bter/enron_hopplot_bter}
    }
\end{minipage}
\begin{minipage}[b]{0.16\linewidth}
    \resizebox{\textwidth}{!}{
        \input{./nfigs/bter/dblp_hopplot_bter}
    }
\end{minipage}
\\
\vspace{-1mm}
\small{BTER}
\vspace{1mm}
\\
\begin{minipage}[b]{0.16\linewidth}
    \resizebox{\textwidth}{!}{
        \input{./nfigs/kron/celegans_hopplot_kron}
    }
\end{minipage}%
\begin{minipage}[b]{0.16\linewidth}
    \resizebox{\textwidth}{!}{
        \input{./nfigs/kron/power_hopplot_kron}
    }
\end{minipage}    
\begin{minipage}[b]{0.16\linewidth}
    \resizebox{\textwidth}{!}{
        \input{./nfigs/kron/arxiv_hopplot_kron}
    }
\end{minipage}%
\begin{minipage}[b]{0.16\linewidth}
    \resizebox{\textwidth}{!}{
        \input{./nfigs/kron/router_hopplot_kron}
    }
\end{minipage}
\begin{minipage}[b]{0.16\linewidth}
    \resizebox{\textwidth}{!}{
        \input{./nfigs/kron/enron_hopplot_kron}
    }
\end{minipage}
\begin{minipage}[b]{0.16\linewidth}
    \resizebox{\textwidth}{!}{
        \input{./nfigs/kron/dblp_hopplot_kron}
    }
\end{minipage}
\\
\vspace{-1mm}
\small{Kronecker}
\\
\begin{minipage}[b]{0.35\linewidth}
    \resizebox{\textwidth}{!}{
        \input{./figs/legend}
    }
\end{minipage}%
\caption{Hop plot. $G$ shown in blue. Results for recurrences $G^\prime_2$, $G^\prime_5$, $G^\prime_8$ and $G^\prime_{10}$ in lighter and lighter shades of red. Degeneration is observed when recurrences increasingly deviate from $G$.}
\label{fig:hop_results}
\end{figure}
\begin{figure}[t]
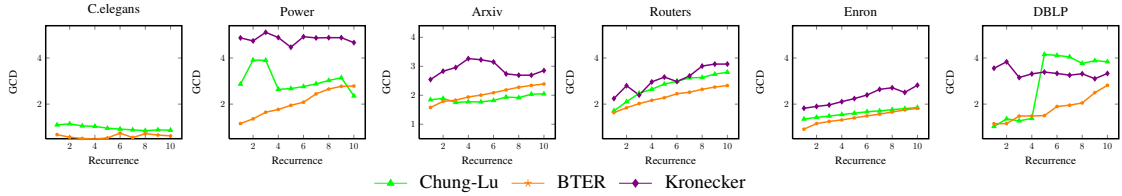

\centering
\begin{minipage}[b]{0.16\linewidth}
    \resizebox{\textwidth}{!}{
        \input{./nfigs/fcl/celegans_gcd_fcl}
    }
\end{minipage}%
\begin{minipage}[b]{0.16\linewidth}
    \resizebox{\textwidth}{!}{
        \input{./nfigs/fcl/power_gcd_fcl}
    }
\end{minipage}    
\begin{minipage}[b]{0.16\linewidth}
    \resizebox{\textwidth}{!}{
        \input{./nfigs/fcl/arxiv_gcd_fcl}
    }
\end{minipage}%
\begin{minipage}[b]{0.16\linewidth}
    \resizebox{\textwidth}{!}{
        \input{./nfigs/fcl/router_gcd_fcl}
    }
\end{minipage}
\begin{minipage}[b]{0.16\linewidth}
    \resizebox{\textwidth}{!}{
        \input{./nfigs/fcl/enron_gcd_fcl}
    }
\end{minipage}
\begin{minipage}[b]{0.16\linewidth}
    \resizebox{\textwidth}{!}{
        \input{./nfigs/fcl/dblp_gcd_fcl}
    }
\end{minipage}
\\
\begin{minipage}[b]{0.35\linewidth}
    \resizebox{\textwidth}{!}{
        \input{./figs/legendnoH}
    }
\end{minipage}%
\caption{Graphlet Correlation Distance. All recurrences are shown for Chung Lu, BTER and Kronecker graph generators. Lower is better. Degeneration is indicated by a rise in the GCD values as the recurrences increase. }
\label{fig:gcd_results}
\end{figure}

Figures~\ref{fig:deg_results},~\ref{fig:eig_results},~\ref{fig:hop_results},~and~\ref{fig:gcd_results} show the results of the Chung-Lu, BTER and Kronecker graphs respectively. 

Different graph generators will model and produce graphs according to their own internal biases. Judging the performance of the generated graphs typically involves comparing various properties of the new graph with the original graph. In Figs.~\ref{fig:deg_results}--\ref{fig:gcd_results} we show the plots of the degree distribution, eigenvector centrality, hop plots and graphlet correction distance. Each subplot shows the original graph in blue and the generated graphs $G^\prime_2$, $G^\prime_5$, $G^\prime_8$, $G^\prime_{10}$ in increasingly lighter shades of red.

In the remainder of this section we will examine the results one metric at a time, \ie, figure-by-figure.

\myparagraph{Degree Distribution.} The degree distribution of a graph is the ordered distribution of the number of edges connecting to a particular vertex. Barab\'{a}si and Albert initially discovered that the degree distribution of many real world graphs follows a heavy-tailed power law distribution such that the number of nodes $N_d\propto d^{-\gamma}$ where $\gamma>0$ and $\gamma$, called the power law exponent, is typically between 2 and 3~\cite{barabasi1999emergence}. 

Figure~\ref{fig:deg_results} shows the degree distribution of Chung Lu, BTER and Kronecker row-by-row for each of the six data sets. The Kronecker generator was unable to model the C. elegans graph because C. elegans does not have a power-law degree distribution, thus those results are absent. These plots are drawn with the original graph $G$ in blue first, then $G^\prime_2$, $G^\prime_5$, $G^\prime_8$ and $G^\prime_{10}$ are overlaid on top in that order; as a result, light-red plots often elide dark-red or blue plots indicating accurate results and non-degeneration. In general, we find that the degree distributions hold mostly steady throughout all 10 recurrences. One exception is present in the Power grid dataset for all three graph generators where the later graphs lose density in the head of their degree distribution. But overall the results are surprising stable.

\myparagraph{Eigenvector Centrality.} The principal eigenvector is often associated with the centrality or ``value'' of each vertex in the network, where high values indicate an important or central vertex and lower values indicate the opposite. A skewed distribution points to a relatively few ``celebrity'' vertices and many common nodes. The principal eigenvector value for each vertex is also closely associated with the PageRank and degree value for each node.

Figure~\ref{fig:eig_results} shows an ordering of nodes based on their eigenvector centrality. Again, the results of Kronecker on C. elegans is absent. With the eigenvector centrality metric we see a clear case of model degeneration in several data sets, but stability in others. The arXiv graph degenerated in Chung-Lu and BTER, but was stable in Kronecker. On the other hand, the Power grid and Routers graph had only a slight degeneration with Chung Lu and BTER models, but severe problems with the Kronecker model.

\myparagraph{Hop Plot.} The hop plot of a graph shows the number of vertex-pairs that are reachable within $x$ hops. The hop plot, therefore, is another way to view how quickly a vertex's neighborhood grows as the number of hops increases. As in related work~\cite{leskovec2005graphs} we generate a hop plot by picking 50 random nodes and perform a breadth first traversal over each graph.

Figure~\ref{fig:eig_results} shows the hop plots of each graph, model and recurrence level. Again we find mixed results. Model degeneration is clear in the arXiv results for Chung Lu and BTER: we see a consistent flattening of the hop plot line recurrence-level increases. Yet the arXiv results are consistent with the Kronecker model. 

The hop plot results are quite surprising in many cases. All of the models severely underestimate the shape of the power grid and routers graphs even in the first generation (not shown).

Of the many topological characteristics that could be compared, researchers and practitioners typically look at a network's \textit{global properties} as in Figs~\ref{fig:deg_results}--\ref{fig:eig_results}. Although these metrics can be valuable, they do not completely test the performance of a graph generator.

In our view, a large network is essentially the combination of many small sub-networks. Recent work has found that the global properties are merely products of a graph's \textit{local properties}, in particular, graphlet distributions~\cite{przulj2007biological}. As a result, graphlet counting~\cite{Marcus2010rage,Ugander2013,AhmedNRD15} and related comparison metrics~\cite{yaverouglu2015proper} comprise the local-side of graph generator performance. 

Thus a complete comparison of graph generator performance ought to include both local and global metrics. In other words, not only should a generated graph have the same degree distribution, hop plot, etc. as the original graph, but the new graph should also have the same number of triangles, squares, 4-cliques, etc. as the original graph. 

There is mounting evidence which argues that the graphlet distribution is the most complete way to measure the similarity between two graphs~\cite{przulj2007biological,Ugander2013}. The graphlet distribution succinctly describes the distribution of small, local substructures that compose the overall graph and therefore more completely represents the details of what a graph ``looks like.'' Furthermore, it is possible for two very dissimilar graphs to have the same degree distributions, hop plots, etc., but it is difficult for two dissimilar graphs to fool a comparison with the graphlet distribution.

\myparagraph{Graphlet Correlation Distance} 

Recent work from systems biology has identified a new metric called the Graphlet Correlation Distance (GCD). Simply put, the GCD computes the distance between two graphlet correlation matrices -- one matrix for each graph~\cite{yaverouglu2015proper}. Because GCD is a distance metric, lower values are better. The GCD can range from $[0,+\infty]$, where the GCD is 0 if the two graphs are isomorphic. 

Figure~\ref{fig:gcd_results} shows the GCD of each recurrence level. Because GCD is a distance, there is no blue line to compare against; instead, we can view degeneracy as an increase in the GCD as the recurrences increase. Again, results from the Kronecker model are absent for C. elegans. As expected, we see that almost all of the models show degeneration on almost all graphs. 

Kronecker's GCD results show that in some cases the GCD is slightly reduced, but in general its graphs deviate dramatically from the original. Chung-Lu and BTER show signs of better network alignment when learning a model from C. elegans. This result highlights biased assumptions in the Chung Lu and BTER models that seem to favor networks of this kind while struggling to handle networks with power-law degree distributions.


\begin{figure}[t]
\centering

\begin{minipage}{.49\textwidth}
\centering

\begin{minipage}[b]{0.42\linewidth}
    \resizebox{\textwidth}{!}{
        \input{./nfigs/fcl/router_cc_fcl}
    }
\end{minipage}%
\begin{minipage}[b]{0.42\linewidth}
    \resizebox{\textwidth}{!}{
        \input{./nfigs/bter/router_cc_bter}
    }
\end{minipage} 
\\
\begin{minipage}[b]{0.42\linewidth}
    \resizebox{\textwidth}{!}{
        \input{./nfigs/kron/router_cc_kron}
    }
\end{minipage}%
\begin{minipage}[b]{0.42\linewidth}
    \resizebox{\textwidth}{!}{
        \input{./nfigs/fcl/router_cc_tcl}
    }
\end{minipage}%
\\
\begin{minipage}[b]{0.42\linewidth}
    \resizebox{\textwidth}{!}{
        \input{./nfigs/fcl/router_cc_fclb}
    }
\end{minipage}    
\begin{minipage}[b]{0.42\linewidth}
    \resizebox{\textwidth}{!}{
        \input{./nfigs/fcl/router_cc_tclb}
    }
\end{minipage}%
\\
\begin{minipage}[b]{0.70\linewidth}
    \resizebox{\textwidth}{!}{
        \input{./figs/legend}
    }
\end{minipage}%
\captionof{figure}{Clustering Coefficient. $G$ is in blue. Results for recurrences $G^\prime_2$, $G^\prime_5$, $G^\prime_8$ and $G^\prime_{10}$ in lighter and lighter shades of red. Degeneration is observed when recurrences increasingly deviate from $G$.} 
\label{fig:cc_results}
\end{minipage}~
\begin{minipage}{.49\textwidth}
\centering

\begin{minipage}[b]{0.42\linewidth}
    \resizebox{\textwidth}{!}{
        \input{./nfigs/fcl/router_assort_fcl}
    }
\end{minipage}%
\begin{minipage}[b]{0.42\linewidth}
    \resizebox{\textwidth}{!}{
        \input{./nfigs/bter/router_assort_bter}
    }
\end{minipage} 
\\
\begin{minipage}[b]{0.42\linewidth}
    \resizebox{\textwidth}{!}{
        \input{./nfigs/kron/router_assort_kron}
    }
\end{minipage}%
\begin{minipage}[b]{0.42\linewidth}
    \resizebox{\textwidth}{!}{
        \input{./nfigs/fcl/router_assort_tcl}
    }
\end{minipage}%
\\
\begin{minipage}[b]{0.42\linewidth}
    \resizebox{\textwidth}{!}{
        \input{./nfigs/fcl/router_assort_fclb}
    }
\end{minipage}    
\begin{minipage}[b]{0.42\linewidth}
    \resizebox{\textwidth}{!}{
        \input{./nfigs/fcl/router_assort_tclb}
    }
\end{minipage}%
\\
\begin{minipage}[b]{0.70\linewidth}
    \resizebox{\textwidth}{!}{
        \input{./figs/legend}
    }
\end{minipage}%
\captionof{figure}{Assortativity. $G$ is in blue. Results for recurrences $G^\prime_2$, $G^\prime_5$, $G^\prime_8$ and $G^\prime_{10}$ in lighter and lighter shades of red. Degeneration is observed when recurrences increasingly deviate from $G$.} 
\label{fig:assort_results}
\end{minipage}

\end{figure}

\myparagraph{Clustering Coefficients.} A node's clustering coefficient is a measure of how well connected a vertex's neighbors are. Specifically, a nodes's clustering coefficient, \ie, the local clustering coefficient, is the number of edges that exist in a node's ego-network divided by the total number of nodes possible in the ego-network. The global clustering coefficient is simply the average of all the local clustering coefficients.

The Chung Lu generator has been shown to model the degree distribution of some input graph, and our results bare this out. Eigenvector centrality, hop plot and graphlet correlation distances are also reasonably well modelled by the Chung Lu generator. However, Pfeiffer \etal~recently showed that the standard Chung Lu generator does not well model a graph's local clustering coefficients; so they introduced the Transitive Chung Lu generator as an adaptation to the standard model~\cite{pfeiffer2012fast}.

\myparagraph{Assortativity.} The assortativity of a network is its tendency to have edges between nodes with similar degree. For example, if high degree nodes primarily link to other high degree nodes, and low degree nodes primarily link to low degree nodes, then the network's overall assortativity score will be high, and vice versa. The local assortativity for each node is the amount, positive or negative, that the node contributes to the overall global assortativity~\cite{newman2003mixing}.

Like in the case with the clustering coefficient, the standard Chung Lu model was found to not accurately model the assortativity of real world graphs. Mussmann \etal~developed a Chung Lu with Binning adaptation that was shown to generate graphs with appropriate assortativity~\cite{mussmann2014assortativity}. Even better is that the transitive and binning models can be combined to create a Transitive Chung Lu with Binning generator that models the degree distribution, clustering coefficient and assortativity of some input graph.

But the question remains, are these new generators robust?

We applied the infinity mirror test to the 6 graph generators, 3 original and 3 Chung Lu adaptations on the Routers dataset. All tests were performed on all graphs for all generators, but cannot all be shown because of space limitations. Figure~\ref{fig:cc_results} shows the clustering coefficient results. We find that transitive Chung Lu does nominally better than standard Chung Lu, but in all cases, the 5th, 8th and 10th recurrences seem to drift away (up and to the right) from original graph's plots demonstrating slight model degeneration as expressed through clustering coefficient. The Kronecker generator did rather poorly in this test. The Kronecker generator didn't seem to have a degeneration pattern, but was simply inconsistent.

The assortativity results are shown in Figure~\ref{fig:assort_results}. We do not see any noticeable improvement in assortativity between the standard Chung Lu and the Chung Lu with Binning generators. We again find that the 5th, 8th and 10th recurrences seem to drift away (downward) from the original graph's assortativity plots demonstrating slight model degeneration as expressed through assortativity. The Kronecker graph also performed poorly on this test, although it is unclear what the nature of the degeneration is.

\nop{
\begin{table*}[h]
\centering
\caption{Basic Statistics}
\renewcommand\arraystretch{1.2}
\begin{tabular*}{\textwidth}{c|rrrr|rr|rr|rr} &
  \multicolumn{4}{c}{Baseline} & 
  \multicolumn{2}{|c}{Chung-Lu} & 
  \multicolumn{2}{|c}{Kronecker}& 
  \multicolumn{2}{|c}{BTER}
  \\\hline
  {Network}    & {$n$}  &{$m$}     &{$z$}  &{$C$} &{$z^*$}&{$C$}  &{$z^*$}&{$C$} &{$z^*$}&{$C$}  \\\hline
  Enron Emails & 36,692 & 183,831  & 10.02 & 0.50 & 10.0  &0.032  &3.66 & 0.00106\\
  ArXiv GR-QC  & 5,242  & 14,496   & 5.53  & 0.53 & 5.54  &0.007  &1.69 & 0.001 \\
  arXiv A-Phys & 18,772 & 198,110  &       & 0.32 & 3.24  &0.010  &3.24&0.00137&  &0.32\\
  Routers      & 6,474  & 13,895   & 4.29  & 0.25 & 4.09  &0.073  & 1.84   & 0.0024\\
  DBLP         & 317,080& 1,049,866& 6.62  & 0.63 & 6.63  &0.00019&6.63 &0.00019 & & \\
  Power grid   & 4941   & 6594     & 2.67  & 0.08 & 2.67  & 0.0007& 0.719& 0.000137\\
  C. elegans neural& 269 & 2965     & 22.04 & 0.41 & 22.16 & 0.186 & - & -\\
  
\end{tabular*}
\label{tab:systematic}
\end{table*}
}

\section{Discussion and Conclusions}

In the present work we introduced the infinity mirror test for graph generator robustness. This test operates by recursively generating a graph and fitting a model to the newly generated graph. A perfect graph generator would have no deviation from the original or ideal graph, however the implicit biases and assumptions that are cooked into the various models are exaggerated by the infinity mirror test allowing for new insights that were not available before.

Although the infinity mirror test shows that certain graph models show degeneration of certain properties in certain circumstances, it is more important to gain insight from how a model is degenerating in order to understand their failures and make improvements. For example, the BTER results in Figs~\ref{fig:deg_results}-\ref{fig:hop_results} shows via the degree, eigenvector and hop plots that the BTER-generated graphs tend to become more spread out, with fewer and fewer cross-graph links, which, in retrospect, seems reasonable because of the siloed way in which BTER computes its model. Conversely, Chung Lu tends to generate graphs with an increasingly well connected core (indicated by the left-skewed hop plots and overestimated eigenvector centrality), but that also have an increasingly large portion of the generated graph that is sparsely connected (indicated by the odd shaped tail in the right-hand side of the eigenvector centrality plots).

A better understanding of how the model degenerates will shed light on the inherent limitations. We hope that researchers and practitioners can consider using this method in order to  understand the biases in their models and therefore create more robust graph generators in the future.

\end{document}